\documentclass[aps,prl,twocolumn,superscriptaddress,letterpaper]{revtex4-2}

\usepackage{graphicx}
\usepackage{bm}
\usepackage{amssymb}
\usepackage{color}
\usepackage{amsmath}
\usepackage{ulem}
\usepackage[colorlinks=true,linkcolor=blue,anchorcolor=black,citecolor=blue,filecolor=black,menucolor=black,runcolor=black,urlcolor=blue]{hyperref}

\begin{document}

\title{Observation of dislocation-induced topological modes in a three-dimensional acoustic topological insulator}

\author{Haoran Xue}
\thanks{These authors contribute equally.}
\affiliation{Division of Physics and Applied Physics, School of Physical and Mathematical Sciences, Nanyang Technological University,
Singapore 637371, Singapore}

\author{Ding Jia}
\thanks{These authors contribute equally.}
\affiliation{Research Center of Fluid Machinery Engineering and Technology, School of Physics and Electronic Engineering, Jiangsu University, Zhenjiang 212013, China}

\author{Yong Ge}
\affiliation{Research Center of Fluid Machinery Engineering and Technology, School of Physics and Electronic Engineering, Jiangsu University, Zhenjiang 212013, China}

\author{Yi-jun Guan}
\affiliation{Research Center of Fluid Machinery Engineering and Technology, School of Physics and Electronic Engineering, Jiangsu University, Zhenjiang 212013, China}

\author{Qiang Wang}
\affiliation{Division of Physics and Applied Physics, School of Physical and Mathematical Sciences, Nanyang Technological University,
Singapore 637371, Singapore}

\author{Shou-qi Yuan}
\affiliation{Research Center of Fluid Machinery Engineering and Technology, School of Physics and Electronic Engineering, Jiangsu University, Zhenjiang 212013, China}

\author{Hong-xiang Sun}
\email{jsdxshx@ujs.edu.cn}
\affiliation{Research Center of Fluid Machinery Engineering and Technology, School of Physics and Electronic Engineering, Jiangsu University, Zhenjiang 212013, China}

\author{Y. D. Chong}
\email{yidong@ntu.edu.sg}
\affiliation{Division of Physics and Applied Physics, School of Physical and Mathematical Sciences, Nanyang Technological University,
Singapore 637371, Singapore}
\affiliation{Centre for Disruptive Photonic Technologies, Nanyang Technological University, Singapore 637371, Singapore}

\author{Baile Zhang}
\email{blzhang@ntu.edu.sg}
\affiliation{Division of Physics and Applied Physics, School of Physical and Mathematical Sciences, Nanyang Technological University,
Singapore 637371, Singapore}
\affiliation{Centre for Disruptive Photonic Technologies, Nanyang Technological University, Singapore 637371, Singapore}

\begin{abstract}
  The interplay between real-space topological lattice defects and the reciprocal-space topology of energy bands can give rise to novel phenomena, such as one-dimensional topological modes bound to screw dislocations in three-dimensional topological insulators. We obtain direct experimental observations of dislocation-induced helical modes in an acoustic analog of a weak three-dimensional topological insulator. The spatial distribution of the helical modes is found through spin-resolved field mapping, and verified numerically by tight-binding and finite-element calculations.  These one-dimensional helical channels can serve as robust waveguides in three-dimensional media.  Our experiment paves the way to studying novel physical modes and functionalities enabled by topological lattice defects in three-dimensional classical topological materials.
\end{abstract}

\maketitle

There are two distinct types of topological phenomena that commonly occur in crystalline materials. The first is band topology defined in reciprocal space, which gives rise to topologically protected boundary modes in topological insulators and related materials \cite{hasan2010, qi2011}. The other involves topological lattice defects defined in real space, which are protected against local perturbations \cite{mermin1979, kosterlitz2017}.  These two types of topological phenomena are usually studied separately, but when they are simultaneously present in a material, their interplay can give rise to many unique phenomena \cite{ran2009, teo2010, imura2011, jurivcic2012, ruegg2013, shiozaki2014, slager2014, sumiyoshi2016, liu2017, queiroz2019, slager2019, wang2020, wang2020b, li2020}.  For example, it has been theoretically predicted that a screw dislocation in a three-dimensional (3D) topological insulator (TI) can host one-dimensional (1D) topologically-protected helical modes \cite{ran2009}. The existence of helical modes is determined by the nonzero value of $\bold{B}\cdot\bold{G_v}/2\pi$, where $\bold{B}$ is the Burgers vector  of the dislocation, defined in real space, and $\bold{G_v}$ is a reciprocal vector defined by the three weak topological indices \cite{fu2007}. This ``bulk-dislocation correspondence'' is a modification of the more well-known bulk-boundary correspondence principle, taking the topology of the lattice dislocation into account. Similar phenomena have been theoretically studied in the context of strong and weak 3D electronic TIs \cite{ran2009, teo2010, imura2011, slager2014}, 3D photonic Chern crystals \cite{lu2018}, and weak 3D photonic TIs with a synthetic dimension \cite{lin2018}.  Dislocation-induced topological modes have technological possibilities as a way to embed robust 1D waveguides in 3D TIs. Compared to other topological waveguide designs such as hinge modes in 3D higher-order TIs \cite{schindler2018, langbehn2017, song2017}, dislocation-induced topological modes can have larger bandwidth (they span the entire bulk gap, unlike hinge modes which usually exist inside a surface gap), and can easily support multimode transport (by increasing the Burgers vector or the spin Chern number).

Thus far, however, there have been few experimental observations of dislocation-induced topological modes. Some signatures of these modes have been found in real materials through conductivity measurements \cite{hamasaki2017} and local density of states measurements \cite{nayak2019}. 
Classical wave systems such as photonic crystals and acoustic crystals have been used to demonstrate localized modes bound to dislocations in 2D photonic Chern insulators \cite{li2018} and weak 2D magnetomechanical TIs \cite{grinberg2020}, disclination waveguiding in 2D valley photonic crystals \cite{wang2020}, and fractional mode density at lattice defects in 2D topological crystalline insulators \cite{peterson2021, liu2021}. However, the topological helical modes induced by a screw dislocation in a 3D TI \cite{ran2009} have not previously been realized.

\begin{figure*}
  \centering
  \includegraphics[width=\textwidth]{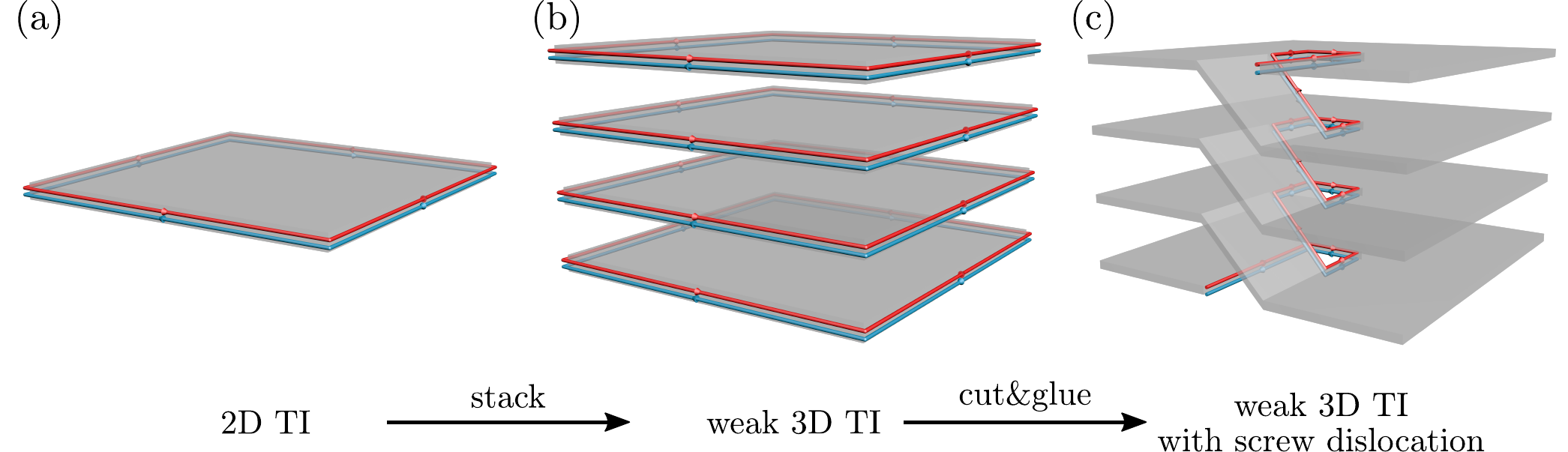}
  \caption{Construction of a 3D weak TI with a screw dislocation. (a) The starting point is a 2D TI with helical edge modes.  (b) To construct a weak 3D TI, the 2D TIs are stacked along $z$.  In the limit of zero inter-layer couplings, the surface modes of the weak 3D TI are nothing but the helical edge modes of the decoupled 2D layers.  (c) A screw dislocation is introduced by a cut-and-glue procedure; the helical edge modes on the bottom and top layers connect to modes running along the dislocation.}
  \label{fig1}
\end{figure*}

In this Letter, we present direct experimental observations of dislocation-induced topological helical modes in a 3D acoustic TI. The sample consists of stacked 2D acoustic TIs implementing a weak 3D acoustic TI with $\bold{G_v}=(0,0,1)\cdot 2\pi/a_z$.  A screw dislocation with $\bold{B}=(0,0,a_z)$ is fabricated into the center of the sample.  With this setup, the bulk-dislocation correspondence predicts a pair of gapless helical modes inside the topological bandgap \cite{ran2009}.  The existence of the helical modes is determined experimentally with field-mapping measurements, resolved over the effective spin and spanning the entire sample, both on the surface and in the bulk.  The experimental results are corroborated by tight-binding calculations and full-wave simulations.

The design process is schematically illustrated in Fig.~\ref{fig1}. We start with a 2D TI that hosts a pair of topologically-protected helical modes on its edges \cite{kane2005, bernevig2006}, as shown in Fig.~\ref{fig1}(a). A weak 3D TI can be constructed by stacking the 2D TIs in the $z$ direction. The bulk bandgap is maximized by letting the couplings between adjacent layers vanish. In that case, the gapless surface states on the side surfaces of the weak 3D TI are intrinsically 2D: they propagate only in the $xy$ plane of each decoupled layer (Fig.~\ref{fig1}(b)).  A screw dislocation along $z$ is then introduced by the ``cut and glue'' procedure shown in Fig.~\ref{fig1}(c).  In tight-binding models as well as the real experimental sample, this can be accomplished by modifying the couplings between certain sites near the center of each layer. The introduction of the dislocation gives rise to a pair of dislocation-induced helical modes, which are connected to the helical edge modes on the bottom and top layers, as shown in Fig.~\ref{fig1}(c).

\begin{figure}[t]
  \centering
  \includegraphics[width=\columnwidth]{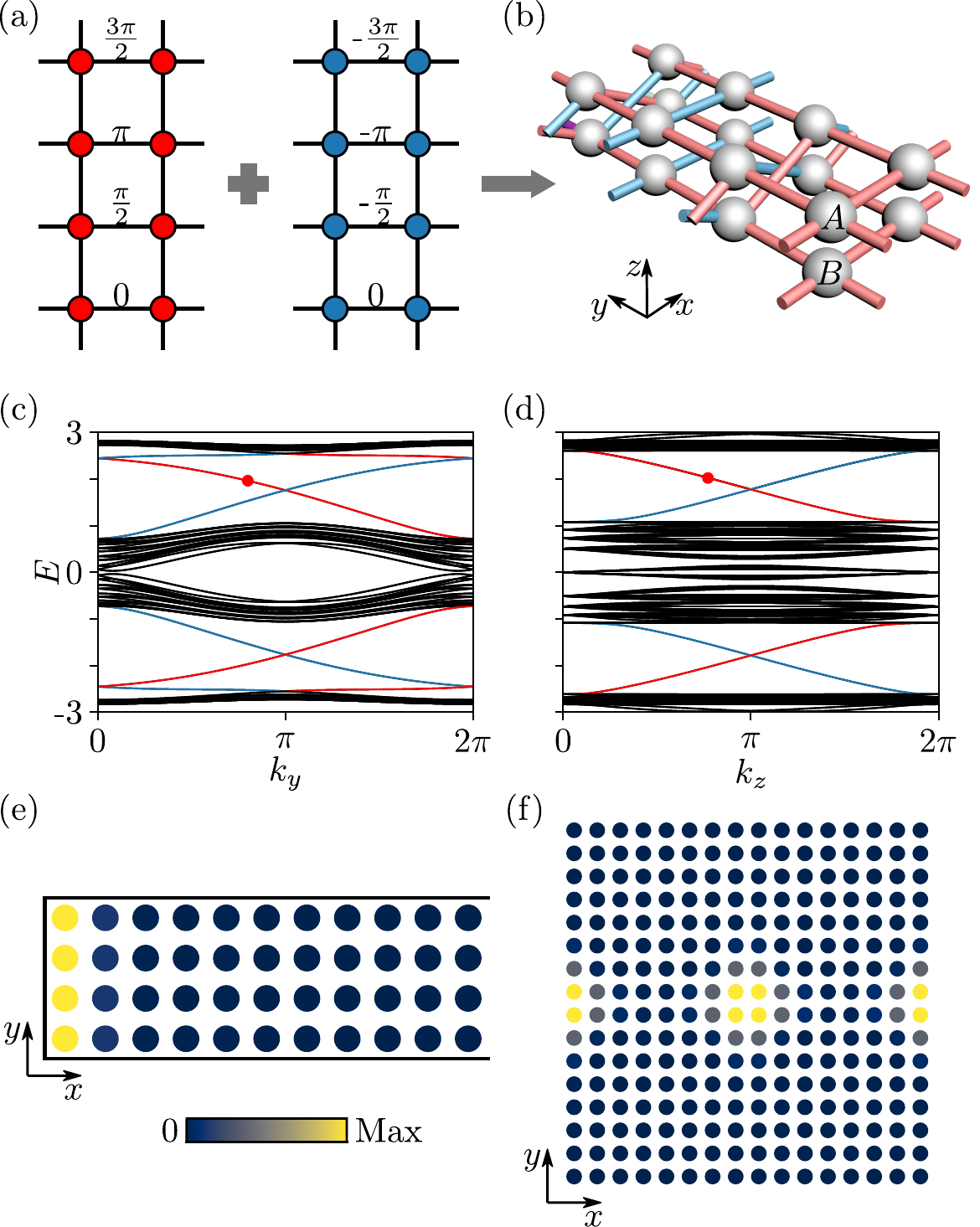}
  \caption{Lattice model design. (a) Square lattices with magnetic flux $\phi=\pi/2$ (left) and $\phi=-\pi/2$ (right) per plaquette. (b) A lattice model for a 2D TI by combining and transforming the two lattice models in (a). Here, all couplings are real and have the same amplitude. Red and blue colors denote positive and negative couplings, respectively. (c) Edge spectrum for the lattice shown in (b), calculated using periodic boundary conditions in $y$ and open boundary conditions in $x$. Black curves correspond to bulk bands and red/blue curves denote spin up/down modes on the left edge. (d) Dislocation spectrum for a weak 3D TI built up by stacking the 2D TI in (b) with vanishing inter-layer couplings. In the calculation, we take a lattice of size 16$\times$16 and use periodic boundary conditions for both $x$ and $y$ directions. The couplings at $x=0$ and $y>0$ are all replaced by inter-layer couplings. This setup leads to two dislocations at (0,0) and (8,0). (e) Profile of the edge mode marked by the red circle in (c). (f) Profile of the dislocation mode marked by the red circle in (d).}
  \label{fig2}
\end{figure}

To design the structure, we begin with a tight-binding model for a 2D TI described by a Hamiltonian of the form
\begin{equation}
H=\begin{bmatrix} H_{\uparrow}(\phi) & 0\\ 0 & H_{\downarrow}(\phi)\end{bmatrix},
\end{equation}
where the submatrices $H_{\uparrow,\downarrow}$ describe 2D Chern insulators consisting of square lattices with complex couplings, as depicted in Fig.~\ref{fig2}(a).  Each is a copy of the Hofstadter model \cite{hofstadter1976}, with magnetic flux $\phi=\pm\pi/2$ (thus $H_{\downarrow}=H_{\uparrow}^{*}$) \cite{hafezi2011, hafezi2013}. The design can be further simplified through the transformation  $[\psi_A,\psi_B]^T=u^{-1}[\psi_{\uparrow},\psi_{\downarrow}]^T$ where \cite{susstrunk2015, ningyuan2015}
\begin{equation}
u=\begin{bmatrix} 1& i\\ 1 & -i \end{bmatrix}/\sqrt{2}.
\end{equation}
Under this transformation, the original complex couplings are replaced by real couplings with either positive or negative signs, as shown in Fig.~\ref{fig2}(b).  A lattice of this sort can be realized in acoustics using acoustic resonators coupled by properly-designed channels \cite{xue2020, ni2020, qi2020}.  Fig.~\ref{fig2}(c) shows the spectrum of the resulting 2D TI, with open boundary condition along $x$ and periodic boundary condition along $y$.  There are two nontrivial bandgaps traversed by helical edge modes, whose mode profiles are shown in Fig.~\ref{fig2}(e).  Finally, following the procedure depicted in Fig.~\ref{fig1}, we stack the 2D TI to form a weak 3D TI, and introduce a central screw dislocation.  This involves replacing the intra-layer couplings along the line segment $y=0$, $x>0$ in each layer (where $(x,y)=(0,0)$ denotes the center of the layer) with inter-layer couplings implementing the ``spiral staircase'' pattern shown in Fig.~\ref{fig1}(c).

Fig.~\ref{fig2}(d) shows the calculated spectrum for a lattice of size 16$\times$16, with periodic boundary conditions imposed along $x$ and $y$ to remove the surface modes on the sides.  With this simulation geometry, there is an extra dislocation at (8,0) in addition to the dislocation at (0,0).  In the plot, dislocation-induced helical modes can clearly be seen within the bulk bandgap.  Each dispersion curve of dislocation modes is two-fold degenerate, corresponding to the two modes at the two dislocations; red (blue) denote spin $\uparrow$ ($\downarrow$) for the modes evaluated at (0,0).  The mode profiles are shown in Fig.~\ref{fig2}(f).

\begin{figure}[t]
  \centering
  \includegraphics[width=\columnwidth]{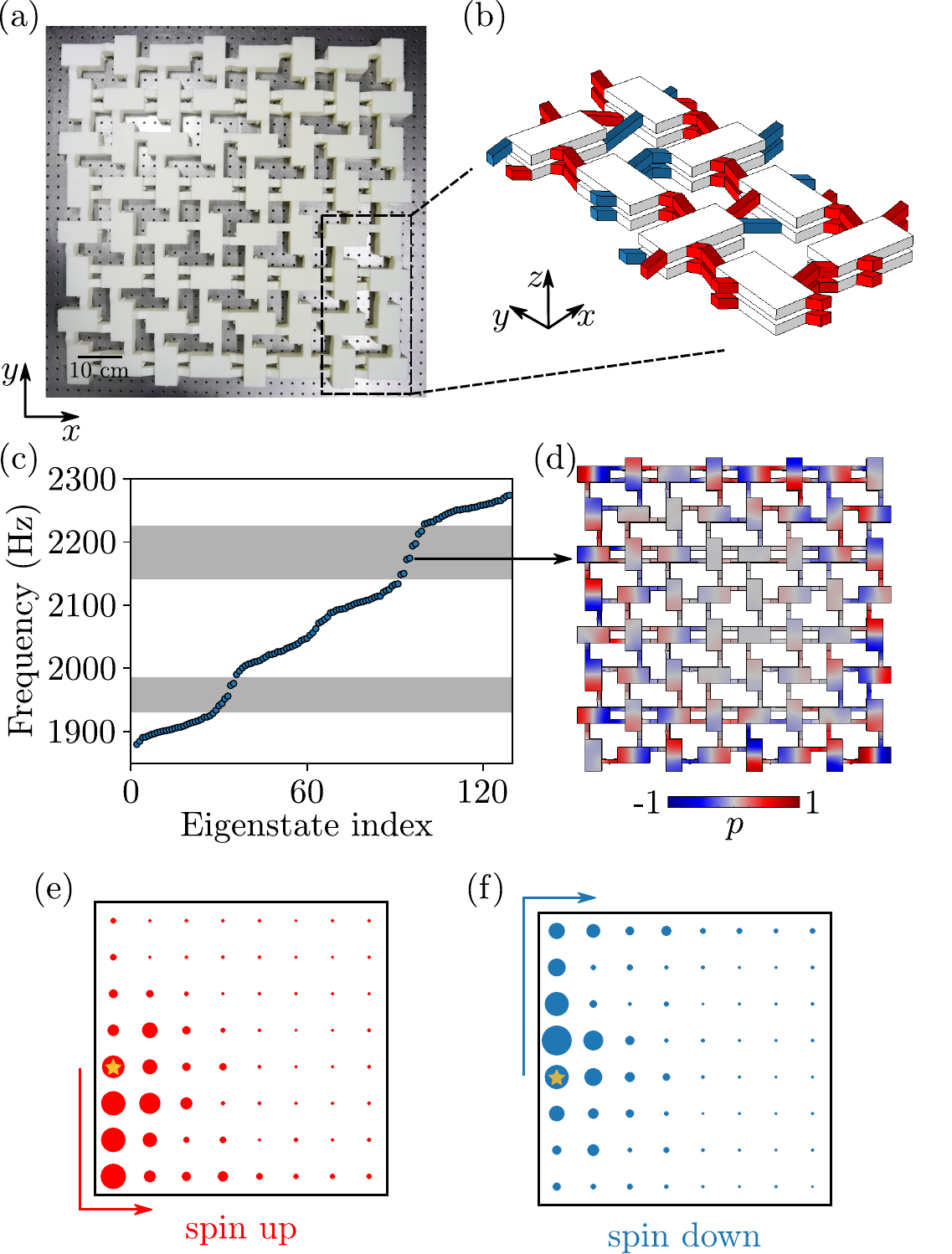}
  \caption{Design and demonstration of a 2D acoustic TI. (a) Photo of the fabricated sample for the 2D acoustic TI. (b) Unit cell for the acoustic lattice, with the red/blue coupling tubes enabling positive/negative couplings. (c) Simulated eigenfrequencies for the acoustic lattice shown in (a). The shaded grey regions denote the bulk bandgap. (d) Eigenmode profile for one of the in-gap edge modes. (e) and (f) Measured field distributions at 2165 Hz for spin up mode (e) and spin down mode (f).}
  \label{fig3}
\end{figure}

\begin{figure*}
  \centering
  \includegraphics[width=\textwidth]{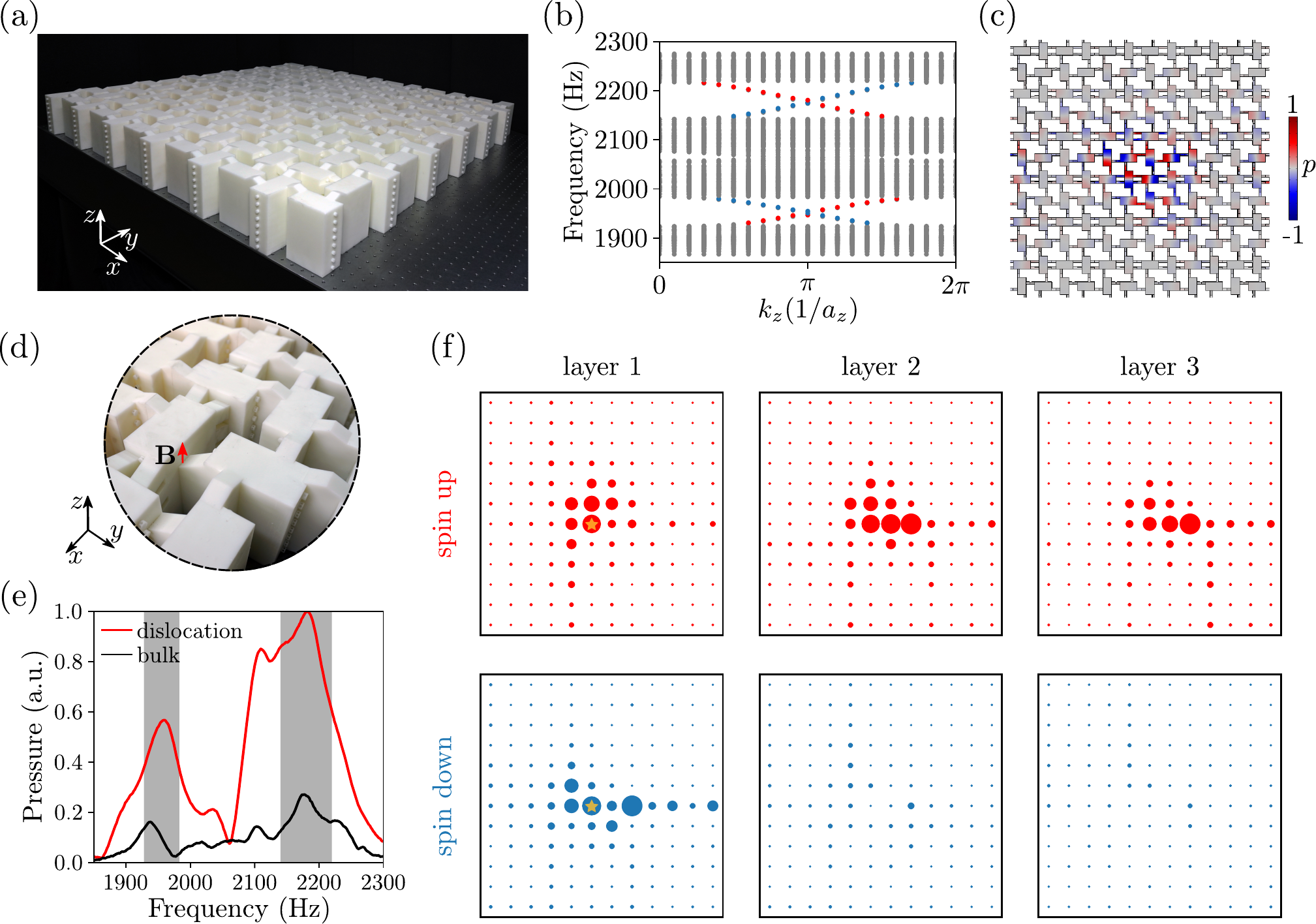}
  \caption{Observation of dislocation-induced topological helical modes. (a) Photo of the fabricated sample for the weak 3D acoustic TI with dislocation. (b) Simulated bandstructure for the lattice shown in (a). Here, grey dots correspond to bulk modes and red/blue dots represent spin up/spin down dislocation modes. (c) Eigenmode profile for one of the dislocation modes. (d) A zoomed-in photo of sample center area. The Burgers vector $\bold{B}$ is denoted by the red arrow. (e) Measured acoustic pressure spectra when the excitation is at the center of the first layer and the detection is at the second layer. Red and black curves are results obtained at one resonator next to the center and away from the center, respectively. (f) Measured acoustic intensity distributions at 2165 Hz in the three layers counted from the top layer. Red and blue colors represent spin up and spin down states, respectively. The subplots within the same column (i.e., the same layer) are normalized by the same value, while the subplots in different columns are normalized by different values.}
  \label{fig4}
\end{figure*}

We now introduce an acoustic lattice that realizes the above tight-binding model.  It consists of two types of building blocks: acoustic resonators and coupling tubes; similar designs have previously been used to create acoustic realizations of other tight-binding models \cite{xiao2015, xiao2017, xue2019a, ni2019, xue2019b, xue2020, weiner2020, ni2020, qi2020}.  A photograph of a one-layer sample (without a dislocation) is shown in Fig.~\ref{fig3}(a), and a schematic of part of it is shown in Fig.~\ref{fig3}(b).  Each cuboid resonator supports a dipolar mode at around 2144 Hz, and is connected to its neighbors through a series of coupling tubes. The coupling tubes shown in red (blue) in Fig.~\ref{fig3}(b) correspond to positive (negative) couplings. Detailed structure parameters of the acoustic lattice are given in the Supplemental Materials \cite{SM}.

Prior to dealing with the full 3D lattice, we verify the features of the 2D lattice using simulations and experiments.  We consider a finite acoustic lattice of size $8\times8$ (Fig.~\ref{fig3}(a)). The simulated eigenfrequencies are plotted in Fig.~\ref{fig3}(c), with the two bulk bandgaps shaded in grey. The modes inside the bandgaps are indeed edge modes, as can be seen from the eigenmode profile plotted in Fig.~\ref{fig3}(d).  Throughout this work, we will focus on the second bandgap at $\sim 2200\,\textrm{Hz}$, which is larger.  

To verify the existence of the edge modes experimentally, we put a speaker at the left edge (yellow star in Figs.~\ref{fig3}(e) and \ref{fig3}(f)) and measure the acoustic field (both amplitude and phase) in each resonator.  Details of the experimental procedure are given in the Supplemental Materials \cite{SM}.  The spin-resolved fields are obtained as $[\psi_{\uparrow},\psi_{\downarrow}]^T=u[\psi_A,\psi_B]^T$. Figs.~\ref{fig3}(e) and \ref{fig3}(f) show the measured intensity distributions at 2165 Hz for spin up and spin down, respectively.  Evidently, the two fields counter-propagate along the edge, as expected of spin-polarized helical edge modes.

Building on this 2D acoustic TI design, we construct a weak 3D acoustic TI with a screw dislocation according to the procedure in Fig.~\ref{fig1}. The fabricated sample is shown in Fig.~\ref{fig4}(a).  It consists of four layers of the 2D acoustic TI with a central screw dislocation, with Burgers vector $\bold{B}=(0,0,a_z)$.  A close-up view of the dislocation is shown in Fig.~\ref{fig4}(d). Fig.~\ref{fig4}(b) shows the simulated dispersion for the 3D acoustic structure, which agrees well with the dispersion obtained using the tight-binding model (Fig.~\ref{fig2}(d)). In particular, we can clearly see the emergence of the dislocation-induced helical modes, with a typical mode profile shown in Fig.~\ref{fig4}(c).

To probe the dislocation-induced helical modes experimentally, we inject sound into one of the resonators at the center of the top layer; the source position is indicated by a yellow star in Fig.~\ref{fig4}(f).  We then scan the acoustic fields inside the whole sample \cite{SM}. Fig.~\ref{fig4}(e) plots the acoustic pressure spectra measured at two different resonators on the second layer (here the second layer refers to the layer right below the top layer). The red curve comes from measurements taken at a resonator next to the dislocation, which has strong peaks at frequencies corresponding to the predicted bulk bandgaps, where the dislocation modes are expected to occur.  The black curve is derived from measurements taken at a resonator that is $2\sqrt{2}a$ away from the dislocation ($a$ is the spacing between adjacent resonators), and has less pronounced peaks at those frequencies. This indicates that the in-gap modes are localized around the dislocation.

Fig.~\ref{fig4}(f) shows site- and spin-resolved acoustic intensity distributions measured at 2165 Hz. As can be seen, the distributions for the two spins are very different. For spin up (upper row), the measured intensity is always localized around the center.  For spin down (lower row), however, the intensities in the second the third layers are negligible. This indicates that only spin up states can propagate downwards along the dislocation; in other words, the direction of propagation along the dislocation is locked to the spin.  The observed spatial profile and helical characteristics of the dislocation-induced topological modes agree well with the simulation results shown in Fig.~\ref{fig4}(b).

To sum up, we have designed and demonstrated a weak 3D acoustic TI with a screw lattice dislocation. The existence of topological helical modes bound to the dislocation is verified both numerically and experimentally. These 1D dislocation-induced helical channels have the potential to serve as robust waveguides in 3D systems. Our results, together with other recent experimental efforts \cite{wang2020b, ye2021}, pave the way to discovering novel phenomena and applications associated with the interaction of nontrivial band topology and topological lattice defects in 3D artificial structures. There are many possibilities for future work, including extending the current design beyond the tight-binding model to achieve a larger bandgap, lower loss and multimode behavior, perhaps by following the ideas in Ref.~\cite{lu2018} but with a non-magnetic design.  Moreover, the introduction of non-Hermiticity, which is available and can be tuned in various artificial topological systems \cite{poli2015, weimann2017, weidemann2020, zhao2019, zhu2018, gao2020, gao2021, helbig2020, liu2020, xiao2020}, may lead to new physical effects yet to be explored \cite{sun2021}. 

H.X., Q.W., Y.C. and B.Z. acknowledge support from Singapore Ministry of Education Academic Research Fund under Grants No. MOE2016-T3-1-006 and MOE2019-T2-2-085. D.J., Y.G., Y.G., S.Y. and H.S. acknowledge support from the National Natural Science Foundation of China under Grants No. 11774137 and 51779107, the China Postdoctoral Science Foundation under Grant No. 2020M671351, and the State Key Laboratory of Acoustics, Chinese Academy of Science under Grant No. SKLA202115.

\end{document}